\def \SAIT #1 #2 {{\em Mem.\ Soc.\ Astron.\ It.\/} {\bf #1}, #2}
\def \MESS #1 #2 {{\em The Messenger\/} {#1}, #2}
\def \ASTRNACH #1 #2 {{ Astron. Nach.\/} { #1}, #2}
\def \AAP #1 #2 {{ A{\rm \&}A\/} {#1}, #2}
\def \AAL #1 #2 {{ A{\rm \&}A\/} {#1}, L#2}
\def \AAR #1 #2 {{ A{\rm \&}AR\/} {#1}, #2}
\def \AAS #1 #2 {{ A{\rm \&}AS\/} {#1}, #2}
\def \AJ #1 #2 {{ AJ\/} {#1}, #2}
\def \ANNREV #1 #2 {{ ARA{\rm \&}A\/} {#1},#2}
\def \APJ #1 #2 {{ ApJ\/} {#1}, #2}
\def \APJL #1 #2 {{ ApJ\/} {#1}, L#2}
\def \APJS #1 #2 {{ ApJS\/} {#1}, #2}
\def \APSS #1 #2 {{ Ap{\rm \&}SS\/} {#1}, #2}
\def \ASR #1 #2 {{ Adv. Space Res.\/} {#1}, #2}
\def \BAIC #1 #2 {{ Bull. Astron. Inst. Czechosl.\/} { #1}, #2}
\def \JSQRT #1 #2 {{ J. Quant. Spectrosc. Radiat. Transfer\/} {
#1}, #2}
\def \MN #1 #2 {{ MNRAS\/} { #1}, #2}
\def \MEM #1 #2 {{ Mem. R. Astr. Soc.\/} { #1}, #2}
\def \PLR #1 #2 {{ Phys. Lett. Rev.\/} { #1}, #2}
\def \PASJ #1 #2 {{ Publ. Astron. Soc. Japan\/} { #1}, #2}
\def \PASP #1 #2 {{ Publ. Astr. Soc. Pacific\/} { #1}, #2}
\def \NAT #1 #2 {{ Nat\/} { #1}, #2}
\def \ACTA #1 #2 {{ Acta Astron.\/} { #1}, #2}
\documentclass[preprint2]{aastex}

\newcommand{\mybold}{}

\slugcomment{To appear in ApJL.}

\shorttitle{IRS spectra of Virgo early type galaxies}
\shortauthors{Bressan et al.}

\usepackage{psfig,latexsym,longtable,lscape,rotating}

\def\smallskip{\vskip 6pt}

\def\M12{${\rm M_{12}}$}

\begin{document}

\title{{SPITZER} IRS spectra of Virgo early type galaxies: \\ 
{\mybold detection of stellar silicate emission.}}
\author{A. Bressan\altaffilmark{1,3,4}, 
P. Panuzzo\altaffilmark{1}, 
L. Buson\altaffilmark{1}, 
M. Clemens\altaffilmark{1}, 
G. L. Granato\altaffilmark{1,4}, 
R. Rampazzo\altaffilmark{1},\\
L. Silva\altaffilmark{2},
J. R. Valdes\altaffilmark{3},
O. Vega\altaffilmark{3},
L. Danese\altaffilmark{4}
} 
\altaffiltext{1}
{INAF Osservatorio Astronomico di Padova, vicolo dell'Osservatorio 5, 35122 Padova, Italy}
\altaffiltext{2}{INAF Osservatorio Astronomico di Trieste, Via Tiepolo 11, I-34131 Trieste, Italy}
\altaffiltext{3}{INAOE, Luis Enrique Erro 1, 72840, Tonantzintla, Puebla, Mexico}
\altaffiltext{4}{SISSA, via Beirut 4, 34014, Trieste, Italy}
\email{ bressan@pd.astro.it; 
panuzzo@pd.astro.it; 
buson@pd.astro.it;
clemens@pd.astro.it;
granato@pd.astro.it;
silva@ts.astro.it;
jrvaldes@inaoep.mx;
ovega@inaoep.mx;
danese@sissa.it}






\date{Received / Accepted }

\begin{abstract}
We present high {\mybold signal to noise ratio} {\it Spitzer} Infrared Spectrograph observations of 17 Virgo early-type galaxies. The galaxies were selected
from those that define the colour-magnitude relation of the cluster, with
the aim of detecting the silicate emission 
of their dusty, mass-losing evolved stars.
To flux calibrate these extended sources we have devised a new procedure
that allows us to obtain the intrinsic spectral energy distribution
and to disentangle resolved and unresolved emission within the same object.
We have found that thirteen objects of the sample (76\%) 
are passively evolving galaxies with a pronounced broad silicate feature
which is spatially extended and likely of stellar origin,
in agreement with model predictions.
The other 4 objects (24\%) are characterized by
different levels of activity. In NGC 4486 (M~87)
the line emission and the broad silicate
emission are evidently unresolved and, given also the typical 
shape of the continuum, they likely originate in the 
nuclear torus. NGC 4636 shows emission lines superimposed on
extended (i.e. stellar) silicate emission, thus pushing the 
percentage of galaxies  with silicate emission to 82\%.
Finally, NGC 4550 and NGC 4435 are characterized by polycyclic aromatic
hydrocarbon (PAH) and line emission,
arising from a central unresolved region. A more detailed analysis of our
sample, with updated models, will be presented in a forthcoming paper.
\end{abstract}



\section{Introduction}
Bressan, Granato \& Silva \cite{Bres98} have
suggested that the presence of dusty circumstellar envelopes
around asymptotic giant branch(AGB) stars should leave a signature,
a clear excess at 10 ${\mu}$m, in the mid infrared (MIR) 
spectral region of passively 
evolving stellar systems. 
Early detections of such an excess were suspected in M32 (Impey et al. 1986) 
from ground based observations, and in a few ellipticals observed with ISOCAM
(Bregman et al. 1998). The first unambiguous confirmation of the existence of
this feature, though barely resolved, was  found in the ISO CVF 
spectrum of NGC 1399 (Bressan et al. 2001). 
Since AGB stars are luminous tracers of intermediate age and  old stellar
populations, an accurate analysis of this feature has been suggested as a
complementary way to
disentangle age and metallicity effects
among early type galaxies (Bressan et al. 1998; 2001).
More specifically, Bressan et al.'s models 
show that a degeneracy between metallicity and age persists even in 
the MIR, since both age and metallicity affect mass-loss 
and evolutionary lifetimes on the AGB. While in the optical
age and metallicity need to be anti-correlated to maintain a 
feature unchanged (either colour or narrow band index),
in the MIR it is the opposite: the larger dust-mass loss
of a higher metallicity {\mybold simple stellar population (SSP)} 
must be balanced by  
its older age. 
Thus a detailed comparison of the MIR 
and optical spectra of passively evolving systems constitutes perhaps one of the cleanest ways
to remove the degeneracy. 
Besides this simple motivation and all other aspects connected
with the detection of evolved mass-losing stars  in passive 
systems (e.g. Athey et al. 2002), 
a deep look into the mid
infrared region may reveal even tiny amounts of activity.
In this letter we present the detection of extended silicate features in
a sample of Virgo cluster early type galaxies, observed with the IRS
instrument\footnote{The IRS was a
collaborative venture between Cornell University and Ball Aerospace
Corporation, funded by NASA through the Jet Propulsion Laboratory and
the Ames Research Center} (Houck et al. 2004) of the 
{\it Spitzer Space Telescope.} (Werner et al. 2004).
\begin{table}
\scriptsize
\caption{Virgo galaxies observed with IRS}
\label{tab1}
\begin{tabular}{lcccccc}
\hline
\hline
Name & V$_T$ & Date  & SL1/2 & LL2  & S/N \\
     &       &       & 60s   & 120s & 6$\mu$m. \\
\hline
NGC~4339  & 11.40  &  Jun 06 2005  & 20 &  14  &      39      \\
NGC~4365  &  9.62  &  Jan 10 2005  &  3 &  3   &      57      \\
NGC~4371  & 10.79  &  Jun 01 2005  &  9 &  10  &      40     \\
NGC~4377  & 11.88  &  Jun 01 2005  & 12 &   8  &      54      \\
NGC~4382  &  9.09  &  Jul 07 2005  &  3 &   3  &      59      \\
NGC~4435  & 10.66  &  Jun 01 2005  &  3 &   5  &      35      \\
NGC~4442  & 10.30  &  Jan 10 2005  &  3 &   3  &      46      \\
NGC~4473  & 10.06  &  Jun 01 2005  &  3 &   3  &      55      \\
NGC~4474  & 11.50  &  Jun 01 2005  & 20 &  14  &      38      \\
NGC~4486  &  8.62  &  Jun 03 2005  &  3 &   3  &      80      \\
NGC~4550  & 11.50  &  Jun 03 2005  & 20 &  14  &      42      \\
NGC~4551  & 11.86  &  Jun 03 2005  & 20 &  14  &      47      \\
NGC~4564  & 11.12  &  Jun 07 2005  &  4 &   6  &      51      \\
NGC~4570  & 10.90  &  Jun 06 2005  &  3 &   5  &      42      \\
NGC~4621  &  9.81  &  Jan 12 2005  &  3 &   3  &      63      \\
NGC~4636  &  9.49  &  Jul 08 2005  &  3 &   5  &      30      \\
NGC~4660  & 11.11  &  Jan 11 2005  &  3 &   5  &      40      \\
\hline
\end{tabular}
\end{table}
\section{Observations and data reduction}
{\mybold Standard Staring mode short SL1 (7.5${\mu}$m-15.3${\mu}$m), 
SL2 (5${\mu}$m-7.6${\mu}$m) 
and long LL2 (14.1${\mu}$m-21.3${\mu}$m), 
low resolution (R$\sim$64-128)}  IRS spectral
observations of 17 early type galaxies, were obtained 
during the first {\em Spitzer} {\mybold General Observer} Cycle.
The galaxies were selected among those that define  
the colour magnitude relation of Virgo cluster (Bower, Lucy \& Ellis 1992).
The observing log is given in Table 1. We also 
report, in columns 4 and 5, the number of {\mybold cycles of 60
and 120 seconds exposures} performed with SL1/2 and LL2, respectively.
The spectra were extracted within a fixed aperture (3\farcs{6}$\times$18" 
for SL) 
and calibrated using custom made
software, tested against the {\tt SMART} software package
(Higdon et al. 2004). 

{\mybold The on-target exposures in each SL segment (e.g. SL1) also provide 
$\sim$80\arcsec\ offset sky spectra in the complementary module (e.g. SL2)
that were used to remove   
the sky background from the source spectrum in the corresponding segment.
Since for the LL module we have obtained only LL2 observations,
LL2 spectra were sky-subtracted by differencing observations in the two nod
positions.}

\subsection{Flux calibration for extended sources}
The IRS pipeline version S12 (and older versions) is designed for
point source flux extraction.
We present here an alternative procedure that exploits
the large degree of symmetry that characterizes 
the light distribution in early type galaxies.

We first obtained the real {\mybold e$^-$/sec to Jy} conversion following the procedure outlined by  Kennicutt et al. (2003). We have corrected the 
conversion table provided for point sources by applying 
the corrections for  aperture losses (ALCF) and slit losses (SLCF).
The ALCF is due to the residual flux falling outside the 
aperture selected in the standard calibration pipeline.
To estimate the ALCF we used 4  calibration stars (HR 2194,
HR 6606, HR 7341 and HR 7891) observed with {\it Spitzer} IRS.
Using the {\it Spitzer} Science Center {\tt SPICE} software package 
we have evaluated the correction resulting
from the average ratio of the fluxes extracted within the standard aperture
and within twice the standard aperture. 
The SLCF correction is applied to retrieve the real
flux of an observed point source that hits the slit,
accounting for the slit losses due to the point spread function
of the optical combination (SLCF).  
It is defined as the wavelength dependent ratio between the whole
flux of a point source on the {\mybold field of view}  and the flux
selected by the slit to hit the detector. To obtain this correction
we have simulated the point spread function of the system
(PSF) using the Spitzer-adapted
{\tt Tiny Tim} code and adopting  a ``hat'' beam transmission
function of the slit. 

After the ALCF and SLCF corrections were applied 
we obtained the flux {\sl received} by the slit within a given aperture. 
The estimate of the flux {\sl emitted} by an extended source within the selected angular
aperture of the slit involves the deconvolution of the received
flux with the PSF of the instrument. This correction is important
to obtain the shape of the intrinsic spectral energy distribution (SED)
of the galaxy, because from the SLCF we have estimated that 
for a point source the losses due to the PSF amount to
about 20\% at 5$\mu$m and to about 40\% at 15$\mu$m. Conversely,
a uniform source will not suffer net losses. 
In order to recover the intrinsic SED we have convolved
a surface brightness profile model with the 
PSF, and we have simulated the corresponding observed linear profile
along the slits, taking into account the relative 
position angles of the slits and the galaxy.
The adopted profile is a wavelength dependent two dimensional
modified King's law (Elson at al. 1987):
\begin{equation}
I \equiv I_0 / \left[1+\frac{X^2}{R_{\rm C}^2}+\frac{Y^2}{(R_{\rm C}\times b/a )^2}\right]^{-\gamma/2}
\label{elson}
\end{equation}
where $X$ and $Y$ are the coordinates along the major and minor axis 
of the galaxies, 
$b/a$ is the axial ratio taken from the literature.
$I_0$, $R_{\rm C}$ and $\gamma$ 
are free parameters that are 
functions of the wavelength and are obtained by fitting the 
observations with the simulated profile. 
In order to get 
an accurate determination of the parameters of the profiles several
wavelength bins have been co-added. 
This procedure has a twofold advantage because
it allows us to (i) reconstruct the intrinsic profile
and the corresponding SED and, (ii)  
to recognise whether a particular feature is resolved 
or not. 
The spectrum, extracted in a fixed width around
the maximum intensity, is corrected by the ratio 
between the intrinsic and observed profile.
Since for the LL2 segment the above procedure is generally not as stable as
for SL segments, we have preferred to fix  $R_{\rm C}$
to the corresponding value derived 
in the nearby wavelength region of the SL segment.

{\mybold 
An estimate of the signal to noise (S/N) ratio was performed by
considering two sources of noise: the instrumental plus background noise and the poissonian noise of the source. The former was evaluated
by measuring the variance of pixel values in background-subtracted coadded
images far from the source.
The poissonian noise of the sources was estimated as the square root of the
ratio between the variance of the number of e$^-$ extracted per pixel in each exposure, and the number of the exposures. 
The total noise was obtained by summing the two sources in quadrature and by multiplying by the square root of the
extraction width in pixels. The corresponding S/N ratio at 6$\mu$m
is shown in column 6 of Table~1. 

Finally we notice that the overall absolute photometric uncertainty
of IRS is 10\%, while the slope deviation within
a single segment (affecting all spectra in the same way)
is less  than 3\% (see the Spitzer Observer Manual). 
}
\begin{figure}
\center{\resizebox{0.49\textwidth}{!}{
\psfig{figure=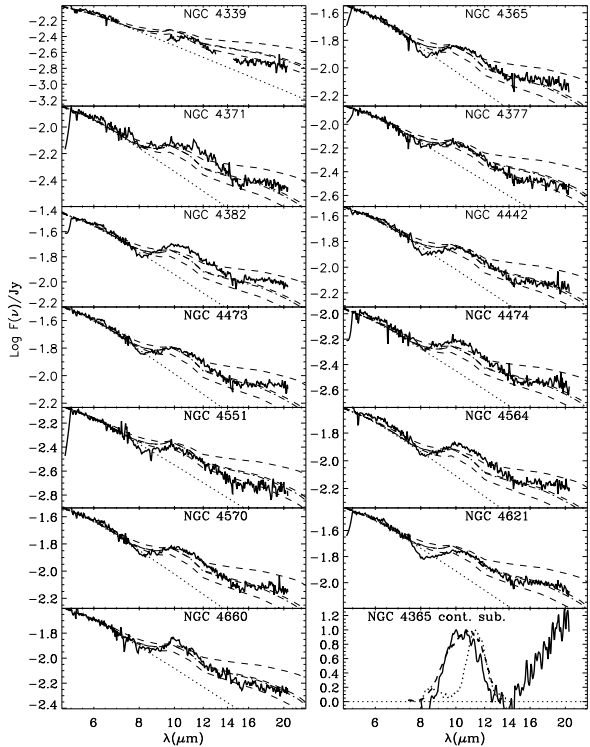,clip=} 
}}
\caption{IRS spectra (solid lines) of
{\sl passively evolving} early-type galaxies in the Virgo cluster.
Superimposed are SSP models from Bressan et al. (1998)
normalized at 5.5$\mu$m. The dotted line is a 10 Gyr, Z=0.02 SSP  computed without accounting for dusty circumstellar envelopes.
Dashed lines from bottom to top  are 10 Gyr SSPs 
with metallicity  Z=0.008, Z=0.02 and Z=0.05 
respectively, computed with dusty circumstellar envelopes.
The dot--dashed line is a young (5 Gyr) metal poor (Z=0.008)
SSP intended to show that also the MIR spectral region suffers
from degeneracy. {\mybold The bottom right panel 
compares the normalized continuum
subtracted spectrum of NGC 4365 with the normalized continuum
subtracted spectrum of the {\sl mean outflow} AGB star
(dashed line, Molster et al. 2002) and the carbon rich star U Cam 
(dotted line, Sloan et al. 1998). 
For NGC 4365 the pseudo-continuum is a straight line that 
interpolates the observed
spectrum between $\lambda$=8$\mu$m and $\lambda$=13.5$\mu$m}.
The spectrum of NGC 4339 is
affected by poorly corrected droops caused by a star falling in the peakup.}
\label{passive}
\end{figure}

\begin{figure}
\plotone{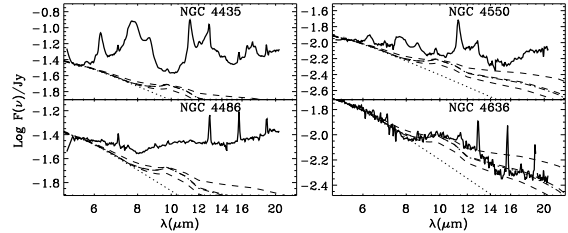}
\caption{IRS spectra (solid lines) of
{\sl active} early-type galaxies in Virgo.
Models are as in Figure \ref{passive}.
}
\label{active}
\end{figure}
\section{Results}
The final flux calibrated spectra 
of the selected Virgo cluster early type galaxies are shown
in Figures \ref{passive} and \ref{active}. 

\begin{figure}
\plotone{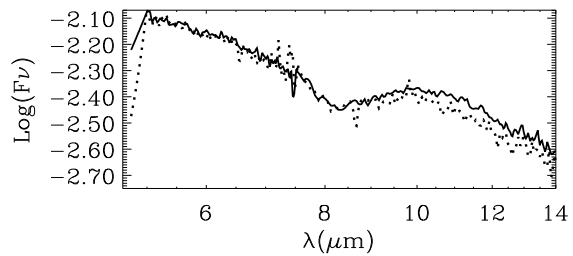}
\caption{{\mybold IRS spectra of NGC 4551 (dotted line) is compared
with that of NGC 4365 (solid line).
Spectra are  normalized at $\lambda$=5.3$\mu$m.}
The silicate features in NGC 4551 are less pronounced 
than in the case of NGC 4365, indicating a smaller dust mass-loss rate
that can be attributed either to an older age or to a lower metallicity.
}
\label{comp}
\end{figure}

\subsection{Silicate emission from evolved stars}
{\mybold In Figure \ref{passive} we have collected  
the thirteen galaxies (76\% of the sample) whose IRS spectra
are characterised by the presence of a broad emission feature 
around $\lambda\sim$10$\mu$m that extends toward longer wavelengths. 
These galaxies show neither PAH features nor emission lines.}  
The observed spectra (solid lines) are superimposed on old SSP from Bressan et al.
(1998) normalized at $\lambda\sim$5.3$\mu$m. The dotted line is a 10 Gyr, Z=0.02 {
\mybold (solar metallicity)} SSP  computed without accounting for dusty
circumstellar envelopes. Dashed lines from bottom to top refer to 10 Gyr SSPs with
increasing metallicity  Z=0.008, Z=0.02 and Z=0.05, computed with dusty
silicate circumstellar envelopes.
{\mybold The models that account for dusty circumstellar envelopes
show an extended feature due to silicate emission, which is
very similar to that observed.
The feature gets stronger at decreasing age and/or at increasing metallicity
due to the corresponding higher dust mass loss rate of the SSP.}
Since, in addition to the match with the models,  
the analysis of the intrinsic spatial profile 
indicates that the whole spectrum is extended, we argue that
the observed features are of stellar origin and
most likely arise from dusty circumstellar envelopes
of mass-losing, evolved stars. 
{\mybold To corroborate this possibility we compare,
in Fig. \ref{passive}, the
normalized continuum--subtracted 10$\mu$m silicate emission 
of the {\sl mean outflow} oxygen-rich AGB star 
(Molster et al. 2002) and  of U Cam (a carbon rich star with SiC emission,
Sloan et al. 1998) with that of NGC 4365.
Though early type galaxies are expected to harbour carbon stars, 
given the wide metallicity spread within a galaxy, 
it seems that the dominant contribution comes from evolved M giants.
More detailed models, fully accounting for the expected mixture
of evolved stars, will be presented in a forthcoming paper.}

The MIR view of early type galaxies proves to be a strong diagnostic for
the population content of these galaxies. Recently  Temi et al. (2005) 
noticed that the IRAC flux ratios at 8$\mu$m and 3.6$\mu$m or MIPS 24$\mu$m
to IRAC 3.6$\mu$m flux ratio remain fairly constant in early type galaxies
which otherwise show different H$\beta$ strength. They conclude that this
disagreement supports a small rejuvenation episode. Although this is one of
the possibilities invoked by Bressan et al. (1996) to explain early type
galaxies with strong Balmer line absorptions, 
caution must be paid before drawing definite conclusions.
Indeed,  we show in Figure \ref{passive} that
a young (5 Gyr) more metal poor (Z=0.008) SSP, dot--dashed line,
is very similar to an old, more
metal rich one. Thus even the mid infrared spectral region is degenerate
and in order to break the age-metallicity degeneracy in
passively evolved systems a careful combined optical ({\mybold including} possibly NIR) and MIR analysis is required.
{\mybold To further illustrate the strength of this kind of analysis,
we show in Figure \ref{comp} a comparison of the IRS spectra 
of NGC 4551 and NGC 4365. 
A recent  optical spectroscopic study (Yamada et al. 2006) indicates
that NGC 4551 is significantly younger and more metal rich 
than NGC 4365. In this case we would expect NGC 4551  to be richer in bright
mass-losing AGB stars than  
NGC 4365, and its silicate features to be more prominent.
However, the opposite is observed, suggesting that
NGC 4551 is either older or more metal poor (or both) than NGC 4365.
Evidently, the effects of degeneracy in the optical can be strong 
(see e.g. Denicol\'o et al. 2005, Annibali et al. 2006).

We finally notice that NGC~4473 was observed by ISO
(Xilouris et al. 2004) and shows 
spatially extended emission at 6.7 and 15$\mu$m.  
These authors measured a 15$\mu$m excess
with respect to SSP models {\sl without} dusty circumstellar  
envelopes. The excess was interpreted as due to hot diffuse 
interstellar dust. 
IRS spectra, such as those presented here, permit the 
disentangling of the
contribution of evolved AGB stars and the presence of interstellar dust.}

\subsection{{\sl Active} galaxies}
The remaining four galaxies (24\% of the sample) display
different signatures of {\it activity} in the MIR spectra (Figure \ref{active}). 
These galaxies are classified as active from optical
studies (from AGN to transition Liner-HII) at odds with the former group.
{\mybold The spectra of NGC 4636 and NGC 4486 (M~87) show emission lines
([ArII]7$\mu$m, [NeII]12.8$\mu$m,  [NeIII]15.5$\mu$m and  [SIII]18.7$\mu$m)
possibly of non-stellar origin. 
The  broad continuum feature at 10$\mu$m in NGC~4486 is not spatially extended and likely due to
silicate emission from the dusty torus (Siebenmorgen et al. 2005; Hao et al. 2005).
Line emission in NGC 4636 falls on top of the circumstellar emission SED
and, as for NGC 4473, its excess at 15$\mu$m is of stellar origin 
and not due to  emission by hot diffuse dust as suggested by
Ferrari et al. (2002).}

The spectrum of  NGC~4550 shows  PAH emissions features
(at 6.2, 7.7, 8.6, 11.3 and 12.7 $\mu$m) and  the H$_2$ S(5) 6.9$\mu$m and 
S(3) 9.66$\mu$m emission lines. 
NGC~4435 shows a typical star-forming spectrum. 
A preliminary interpretation suggests that an unresolved starburst is
dominating the MIR emission (Panuzzo et al. in preparation).

\section{Conclusions}
We presented {\it Spitzer} MIR IRS spectra of early type 
galaxies selected along the colour-magnitude relation of the Virgo cluster. 

{\mybold We have reconstructed the intrinsic SED of these galaxies
from the observed spatial profile sampled by the slits,
via  a careful analysis of PSF effects.
In this way we are also able to differentiate between
spatially resolved and unresolved regions within the spectrum.
This provides independent support for the interpretation of their nature.}

Most of the galaxies (76\%) show an
excess at 10$\mu$m and longward which appears spatially
extended and is likely due to silicate emission.
This class of spectra do not show any other emission features.
We argue that the 10$\mu$m excess  arises 
from mass-losing evolved stars, as predicted by adequate SSP
models. A detailed modelling of these features 
together with the analysis of combined optical, NIR and MIR spectra
will be presented in a forthcoming paper.

In the remaining smaller fraction (24\%) we detect signatures of {\it activity}
at different levels. We observe line emission superimposed on the stellar
silicate features in NGC 4636, unresolved line and silicate emission in
M~87 that likely originate in the dusty torus and unresolved PAH emission
in NGC 4550 and NGC 4435. The latter galaxy displays the
main characteristics of a nuclear starburst (Panuzzo et al. in preparation). 

If we exclude M~87, which is a well known AGN, only two out of 16 early-type
galaxies observed show PAHs, which corresponds to quite a low fraction 
($\sim$12\%) of the observed sample. It is premature to conclude that such a
low fraction of galaxies with PAHs is representative of the cluster 
early-type galaxy population, especially if we consider that our investigation
is limited to the brightest cluster members (the upper two magnitudes
of the colour-magnitude relation). A detailed comparison of our results with
those obtained for field galaxies will cast light on the role of
environment in the galaxy evolution process.

\begin{acknowledgements}
This work is based on observations made with the Spitzer Space Telescope, which
is operated by the JPL, Caltech
under a contract with NASA. We thank J.D.T. Smith for helpful suggestions on the
IRS flux calibration procedure and the anonymous referee for useful suggestions.
A. B., G.L. G. and L. S.  thank INAOE for warm hospitality.
\end{acknowledgements}

\end{document}